\documentclass[runningheads,citeauthoryear]{DoklBAN}
\usepackage{epsfig,cite,graphics,graphicx} 
\usepackage[utf8]{inputenc}
\usepackage{xcolor}

\begin{document}
\title{{\color{blue} {\small Comptes rendus de l'Academie bulgare des Sciences} \\
{\small Tome 79, No 3, 2026  \\ 
$\;$  \\
$\;$ https://doi.org/10.7546/CRABS.2026.03.01 \hskip 4.9cm  $\;$ \\
$\;$ \hskip 9.8cm  ASTRONOMY}} 
\\ 
$\;$ \\   $\;$ \\
BV photometry of the ultracompact binary star GP~Com} 
\titlerunning{ GP Com - a binary white dwarfs system} 
\author{Radoslav Zamanov,$^1$
        Lyuba Dankova,$^1$
        Milen Minev,$^1$ 
        Daniela Boneva,$^2$ 
        Krasimira Yankova$^2$ 
	\vskip 0.3cm 
	$ \; $
	}
\authorrunning{Zamanov, Dankova, Minev et al.}

\papertype{{\it Presented  by N. Nedyalkov, Corresponding Member of BAS, on February 24, 2026}}
\maketitle

\begin{abstract}
We present optical B and V band photometry of GP Com --
an ultracompact binary consisting of an accreting white dwarf and helium secondary
component. 
Our data set contains 7.7 hours observations in V band with the 2.0m telescope 
and 2.9 hours simultaneous observations in B and V bands with the 1.5m telescope
of the Rozhen National Astronomical  Obsevatory, Bulgaria. 
The observations cover  $\approx 13$ orbital periods. 

We find  an orbital modulation with amplitude $0.04 - 0.05$~mag in B and V bands.
Adopting that it is due to a bright spot, we estimate its
temperature $\approx 19700 \pm 3000$~K.
We estimate mass accretion rate onto the white dwarf  of about $2 \times 10^{-12}$~M$_\odot$~yr$^{-1}$,
consistent with the predicted rate for a cool donor. 

The data are available on Zenodo:  zenodo.org/records/18768211. 
\end{abstract}

\section{Introduction}

GP Com (G~61-29) is a $16 th$ magnitude star in the constellation  Coma Berenices
located at a distance 72.7 pc from the Earth$[^1]$. 
It has orbital period$[^2]$  $P_{orb} = 46.57$~min and 
is one of the brightest members of the group of
AM Canum Venaticorum (AM~CVn) stars.  

AM~CVns are binaries with short orbital periods, 
in which the primary star is a
white dwarf accreting from a He-rich secondary.  
They are subclass of the cataclysmic variables family 
and have spectra  rich in helium-emission lines.
AM~CVns are sometimes named interacting binary white dwarfs
and are expected to be sources of persistent gravitational wave radiation.
Their orbital periods range between 5 and 65 minutes$[^3,^4]$
and till now of about 120 objects are found and confirmed$[^5]$. 

Here, we report optical photometric observations of GP~Com 
in Johnson B- and V-bands, study the orbital modulation and estimate 
the temperature of the hot spot and the mass accretion rate.

\begin{figure}
\includegraphics[width=9.8cm]{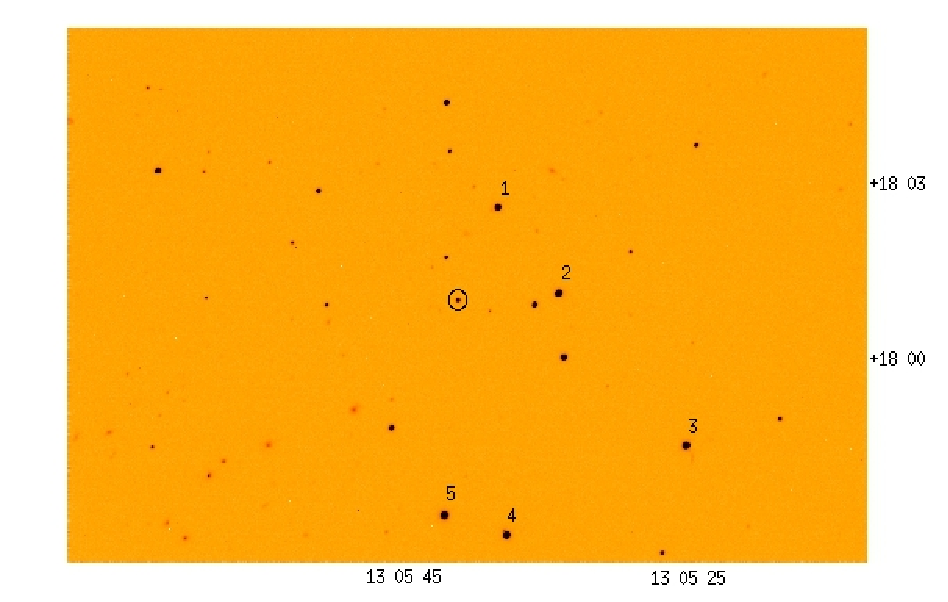}  
\caption[]{V band image of the field around GP~Com, obtained with the 1.5m 
     telescope of NAO Rozhen on 2025-11-14 UT~03:24, exposure time 60 sec. 
    GP~Com is marked with circle. The coordinates and magnitudes of 
    the comparison  stars  can be found in Section~2  }
\end{figure}

\section{Observations} 
\label{s.obs}
The observations were secured with the 2m RCC telescope 
and with the 1.5m AZ1500 telescope$[^6]$ 
of the Rozhen National Astronomical Observatory, Bulgaria. 
Both telescopes are  equipped with  CCD cameras.  
The field of view of the 2m telescope is a circle with diameter 14 arcmin, 
of the 1.5m telescope -- a rectangle  $12 \times 8$ arcmin. 
As a first step we identified a few stars in the field (see Fig.~1)
that can be used as comparison stars: \\

{\small
\begin{tabular}{l  | l l l  | c  c | c  c}
 No. &  RA	    & & Declination    & B	   &	V    & \\  
\# 1 &  13 05 39.4  & &      +18 02 45 & 15.170    &  14.249 & \\  
\# 2 &  13 05 34.6  & &      +18 01 21 & 14.66$^*$ &  13.762 & \\  
\# 3 &  13 05 24.5  & &      +17 58 54 & 14.155    &  13.399 & \\  
\# 4 &  13 05 37.0  & &      +17 57 07 & 14.623    &  13.699 & \\  
\# 5 &  13 05 41.6  & &      +17 57 22 & 13.566    &  13.039 & \\  
\\
\end{tabular}
} 
\\
All they are located at angular distance less than 5 arcmin from GP~Com.
The B and V  magnitudes are synthetic photometry
from GAIA catalog - Vizier catalog I/360/syntphot$[^7]$. 
Their typical errors are 0.002~mag.  The exception is marked with $^*$ and is our measurement.
Good agreement ($\approx 0.01$~mag) between the GAIA magnitudes 
and our measurements was achieved  for stars \#1, \#3, \#4, \#5.  
Using these comparison stars, we measure the B and V magnitudes of GP~Com. 
Journal of observations and the measurements are  presented in Table~1. 

\begin{table*}[ht!]
\caption{Journal of observations and measured magnitudes. }
 \begin{center}
 \begin{tabular}{l | c  l  | c | cc ccr cccll lll}
 \hline
 date       & telescope & band &   UT	        & $N_{pts}$ & exposure & \\
            &           &      & 	        &	    &  [sec]   & \\
	    &           &      &                &           &          & \\  
 2025-04-29 & 2.0m      & V    & 20:29 -- 22:08 &  22	    &   180    & \\ 
 2025-04-30 & 2.0m      & V    & 18:52 -- 00:57 & 116	    &   180    & \\
	    &           &      &                &           &          & \\
 2025-11-13 & 1.5m      & B    & 02:42 -- 03:48 &  30       &    60    & \\
	    &           & V    &		&  30       &    60    & \\
 2025-11-14 & 1.5m      & B    & 02:51 -- 03:55 &  30       &    60    & \\
	    &           & V    &		&  30       &    60    & \\
 2025-11-15 & 1.5m      & B    & 02:42 -- 03:28 &  21       &    60    & \\
	    &           & V    &                &  21       &    60    & \\
	    &           &      &                &           &          & \\  
 \hline                                                                                   
 \end{tabular}                                                                                    
\begin{tabular}{l | c | c c c c c c c r }
 \hline
   date     & band & min    & max    & average & stdev & amplitude &  err  & \\  
            &      & [mag]  & [mag]  & [mag]   & [mag] & [mag]     & [mag] & \\
            &      &        &        &         &       &           &       & \\ 
 2025-04-29 &  V   & 16.130 & 16.187 & 16.1645 & 0.013 & 0.057     & 0.007 & \\ 
 2025-04-30 &  V   & 16.138 & 16.200 & 16.1696 & 0.013 & 0.062     & 0.007 & \\
 2025-11-13 &  B   & 16.051 & 16.118 & 16.0771 & 0.016 & 0.067     & 0.010 & \\
            &  V   & 16.061 & 16.105 & 16.0799 & 0.011 & 0.044     & 0.009 & \\
 2025-11-14 &  B   & 16.024 & 16.078 & 16.0580 & 0.015 & 0.054     & 0.008 & \\	    
            &  V   & 16.032 & 16.094 & 16.0664 & 0.016 & 0.062     & 0.007 & \\
 2025-11-15 &  B   & 16.049 & 16.117 & 16.0753 & 0.013 & 0.068     & 0.008 & \\
            &  V   & 16.066 & 16.102 & 16.0807 & 0.010 & 0.036     & 0.007 & \\ 
            &      &        &        &         &       &           &       & \\ 
\hline  
 \end{tabular}                                                                                  
 \end{center} 
 \label{t.j}                                       
 \end{table*}                                                                                       


\section{Results}
The time-resolved spectroscopy
with VLT/ESO $[^8]$ gives for GP Com
orbital ephemeris  HJD = 2452372.5994(2) + 0.0323386E,  
mass of  the primary component  $M_1 > 0.33$~M$_\odot$ and 
mass of the secondary in the range $9.6 < M_2 < 42.8$~M$_{Jupiter}$
($0.01 - 0.04$~M$_\odot$). 
The orbit is inclined to the line of sight
at an inclination angle  $33^\circ  < i < 78^\circ$.  

\begin{figure}
  \vspace{3.8cm}   
  \includegraphics{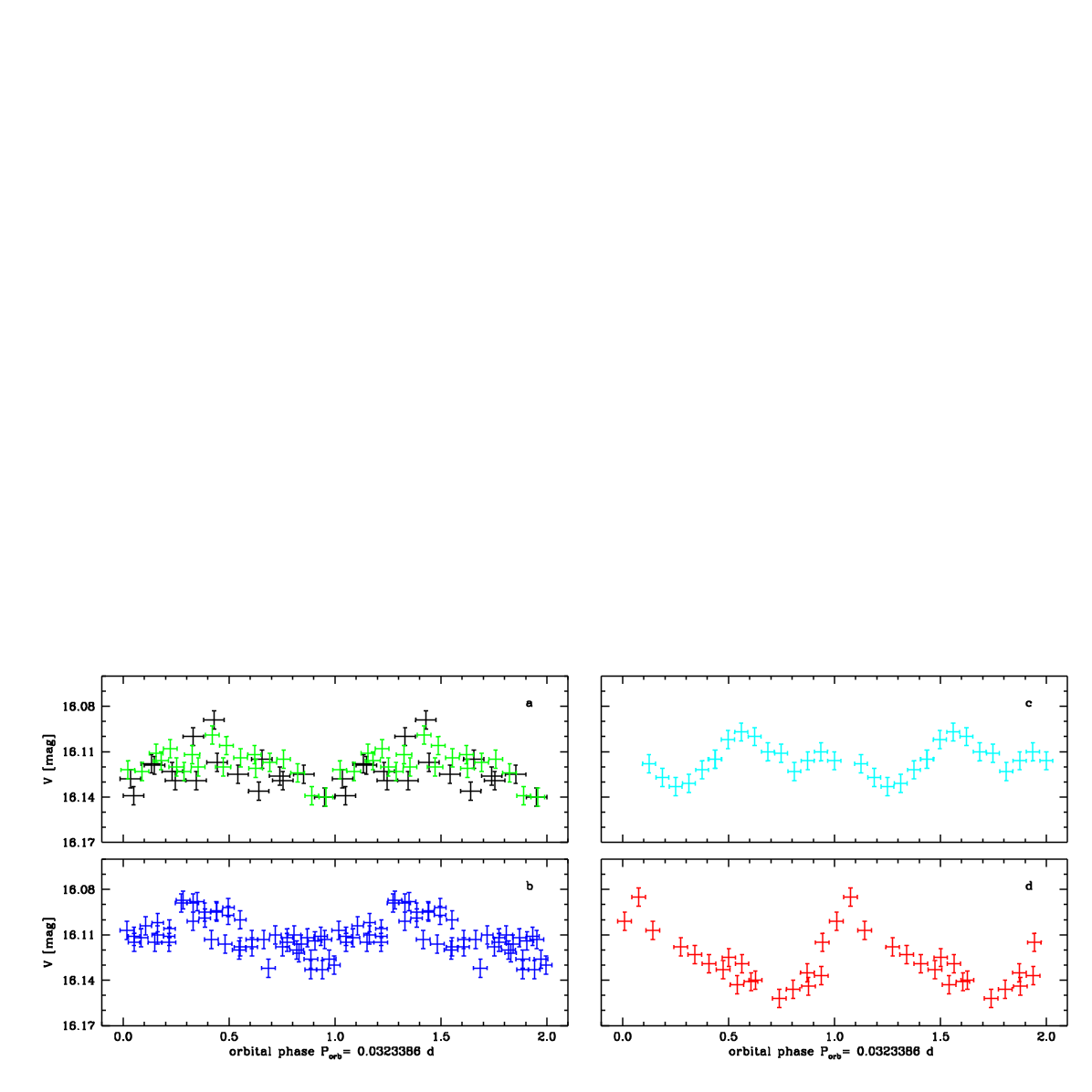} 
  \caption[]{Orbital variability of GP~Com in V band observed with the 2.0m telescope:  
  a)~black colour marks the  observations on 2025-04-29 from UT20:42 to 22:05,
    green - those on 2025-04-30 from UT18:52 to 19:59;
  b)~blue colour - observations on 2025-04-30 from UT20:02 to 22:34;
  c)~cyan - 2025-04-30 from UT22:57 to 23:44;  
  d)~red -  2025-04-30 from UT23:47 to 00:53.
  It is visible that the amplitude of the orbital variability 
  is $\sim 0.04$~mag and the phase of the peak is variable.
  }
 \vspace{7.9cm}  
 \includegraphics{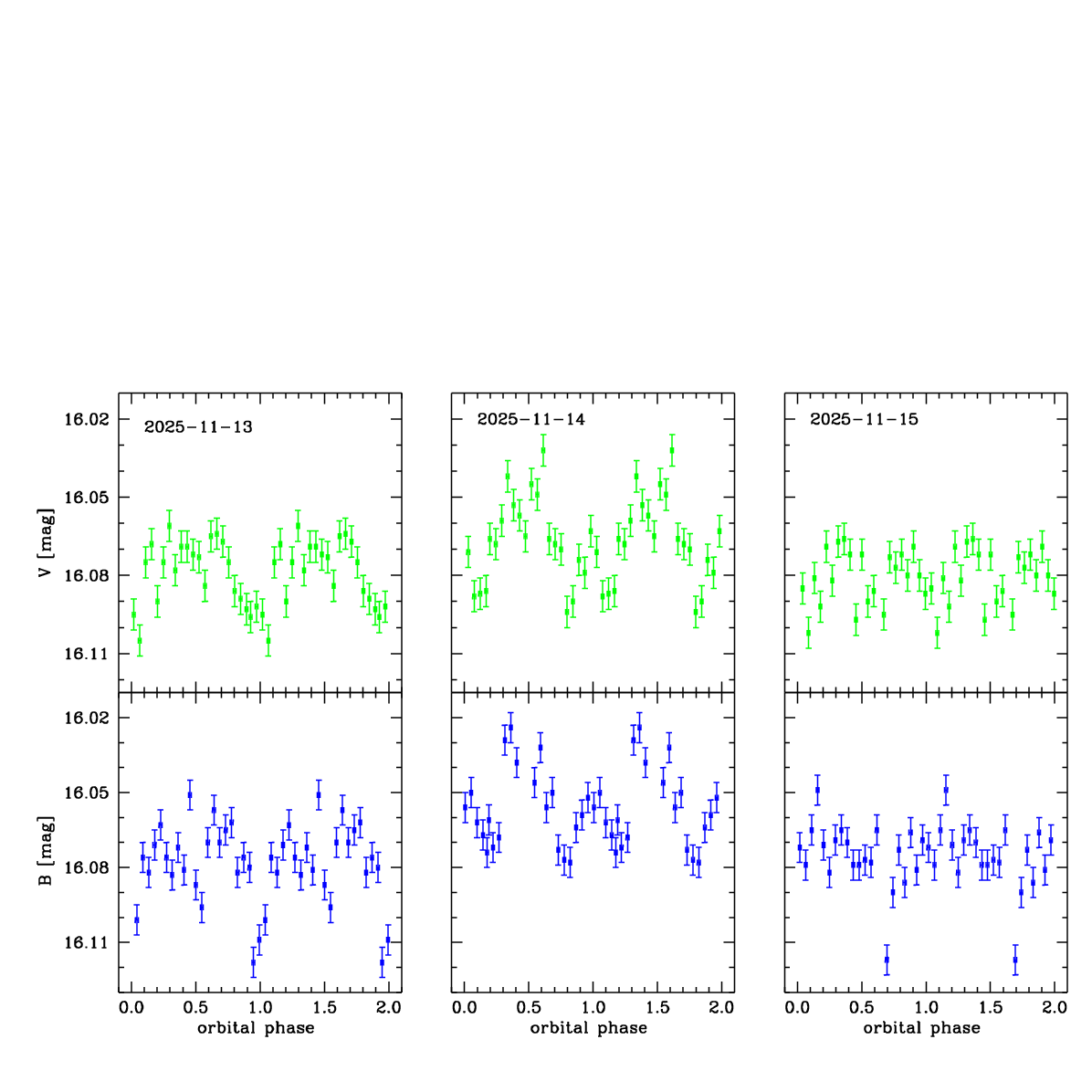} 
 \caption[]{Simultaneous B- and V-band observations of GP~Com
 obtained with the 1.5m telescope. 
 During the first two nights (2025-11-13 and 2025-11-14) 
 orbital variability with amplitude $\approx 0.04$~mag 
 is visible in both bands. } 
\end{figure}

\subsection{Orbital modulation in V band}
\label{s.or}

Using this ephemeris, we plot the orbital variability 
in Fig.2 and Fig.3. 
The observations obtained in April 2025 with the 2.0m telescope
are plotted in Fig.2. They are separated in 4 panels, because 
the orbital variability is not same during the different orbital cycles. 
It is visible that the amplitude of the orbital variability 
is $0.04 -0.05$~mag and the phase of the peak is variable.

In Fig.2a, the black colour marks the  observations on 2025-04-29 from UT20:42 to 22:05,
and the green colour - those on 2025-04-30 from UT18:52 to 19:59. 
Although the observations cover 3 orbital cycles from two different nights
they have very similar orbital modulation with maximum at orbital phase $\sim 0.4$.
In Fig.2b (blue colour) are plotted  the observations 
on 2025-04-30 from UT20:02 to 22:34 --
the peak is at phase $\sim 0.3$.
In Fig.2c (cyan colour) are plotted the  observations 
on  2025-04-30 from UT22:57 to 23:44. The peak is at  phase $\sim 0.5$ 
and a secondary peak at phase $\sim 1.0$ emerges. 
In Fig.2d (red colour) is plotted the next orbital cycle
(2025-04-30 from UT23:47 to 00:53) -- the peak is at phase $\sim 1.0$.

The observations  obtained with the 1.5m telescope are presented in Fig.3. 
During the first two nights (2025-11-13 and 2025-11-14) 
orbital variability with amplitude $\approx 0.04$~mag 
is visible in  B and V bands. During the third night 
no orbital variability can be seen, probably because the run is shorter. 

Our observations show that in the most cases (Fig.2a,b and Fig.3)
the maximum of the orbital variability is at orbital phase 0.3-0.5.
However on Fig.2c a secondary maximum at phase 1.0 emerges 
and becomes the only maximum on Fig.2d. This can be due to
azimuthal displacement of the hot spot or appearance of 
secondary spot (see Sect.\ref{s.dis}).

\subsection{Binary parameters: is the secondary a white dwarf?}

The mass ratio of the components in GP~Com is  $q \approx M_2 /M_1 \approx 0.025$$[^8]$.  
Adopting their value$[^8]$ for the mass of the secondary $M_2 \approx 0.02$~$M_\odot$, 
it will give mass of the primary $M_1 \approx 0.8$~$M_\odot$. 
A  well known  property of the white dwarfs is that the more massive white dwarfs 
are smaller.  To estimate the radius of the white dwarf, 
we use the equation by P. Eggleton$[^{10}]$. 
It gives for the radius of the primary $R_1 = 6970$~km (0.01~$R_\odot$).

The distance between the components, $a$, is 
related to the orbital  period by the Kepler's third law:
$  P_{orb}^2 =  4 \: \pi^2 \: a^3 \: [ G \: (M_1 + M_{2})]^{-1} $.
It gives  $a \approx 0.40$~$R_\odot$. 
Using the formula$[^{11}]$:  
$R_{L2} /a  =  0.49 \: q^{2/3} [ 0.6 q^{2/3} + \ln (1+q^{1/3}) ]^{-1} $
and mass ratio
$0.025$, we estimate Roche lobe size of
the primary   $R_{L1} \approx 0.269$~$R_\odot$
and of  
the secondary $R_{L2} \approx 0.054$~$R_\odot$. 
The accuracy of our estimates is connected with the uncertainty of 
the binary  parameters$[^8]$, and is of about 5-10\%.  
 
The mass transfer in  AM~CVn stars as well as in cataclysmic variables in general, 
is via Roche lobe overflow. 
When the mass transfer has already started, 
the Roche lobe size of the secondary is approximately equal to its radius.
In other words the radius of the secondary is $R_2 \approx R_{L2} \approx 0.054$~$R_\odot$.
However a white dwarf with mass 0.025~$M_\odot$
should have a radius $\approx 0.039$~$R_\odot$, which is 1.4 times  less than 
the estimated Roche lobe size. This result indicates 
that the secondary in GP~Com is not a white dwarf, but probably
a helium star (the helium core of a former main sequence star).

\subsection{Absolute magnitude and mass accretion rate}
\label{s.Ma}

GP~Com  has coordinates J2000 RA~13:05:42.4,  Dec~+18:01:03, 
corresponding to galactic coordinates 324$^\circ$~+80$^\circ$.
The object is located in  direction close to the  north galactic pole.
In this direction there is no interstellar dust in our Galaxy, 
and the interstellar extinction to GP~Com is zero$[^9]$. 

Combining our data $16.02 \le V \le 16.20$ 
and  $16.02 \le B \le 16.12$ (see Table~1)
with the GAIA$[^1]$ distance (72.7~pc), we find for GP~Com 
absolute V band magnitude in the range  $11.7 <  M_V < 11.9$ 
and  absolute B band magnitude in the range $11.7 <  M_B < 11.8$.
Our values are in agreement with the result$[^{12}]$,
that GP~Com is intrinsically faint, $M_V > 11$.

From the three nights simultaneous B and V band observations with the 1.5m telescope,
we find that the maximum brightness is 
$B=16.042 \pm 0.014$, $V=16.053 \pm 0.018$, $B-V = -0.011$. 
This corresponds to a  black body 
(located at distance 72.7~pc from the Earth)
with effective temperature $T=13200 \pm 200$~K, 
effective radius $R=0.011 \pm 0.001$~$R_\odot$,
and luminosity $L \approx 1.21 \times 10^{31}$~erg~s$^{-1}$ (0.0031~L$_\odot$).
The optical luminosity of an accretion disc 
is connected with the mass accretion rate: 
$L = 0.5 \: G \;  M_{1} \;  \dot M_a \: R_{1}^{-1}$,
where $G$ is the gravitational constant, 
$\dot M_a$ is the mass accretion rate,
$R_{1}$ is the radius of the white dwarf. 
For a standard accretion disc, the disc luminosity 
is half of the total accretion luminosity.
We calculate  mass accretion rate of about 
$\dot M_a = 4.2 \times 10^{-12}$~M$_\odot$~yr$^{-1}$
and 
$ 2.5 \times 10^{-12}$~M$_\odot$~yr$^{-1}$
for $M_1 =0.6$  and 0.8~M$_\odot$, respectively. 
These values should be accurate to within a factor of two.

For GP~Com a mass accretion rate
$\dot M_a = 2.3  \times 10^{-11}$~M$_\odot$~yr$^{-1}$
is estimated$[^{13}]$.
Our results indicate 6 to 9 times lower mass accretion rate
and are  consistent 
with the predicted mass accretion rate for a cool donor (see Fig.4 in $[^{13}]$).

\subsection{Orbital modulation: temperature of the hot spot}

In cataclysmic variables 
the observed luminosity is a sum of the luminosity of 
the accreting white dwarf, accretion disc, mass donor, and a hot spot. 
The mass is transferred in a stream through the inner Lagrangian point 
to an accretion disc. 
At  the point where the stream impacts the disc, a hot spot is formed.
The orbital variability is a result of the hot spot -- 
the maximum is when the hot spot faces to the observer.  

To calculate the temperature of the hot spot, we first 
convert the observed magnitudes in fluxes, using the calibration 
for a zero magnitude star 
$F_0 (B) = 6.13268 \times 10^{-9}$  erg cm$^{-2}$ s$^{-1}$ \AA$^{-1}$,     $\lambda_{eff}(B)=4371.07$~\AA, 
$F_0 (V) = 3.62708 \times 10^{-9}$  erg cm$^{-2}$ s$^{-1}$ \AA$^{-1}$ and  $\lambda_{eff}(V)=5477.70$~\AA\ 
as given in the Spanish virtual observatory 
Filter Profile Service[$^{14}$].   
Adopting that the orbital variability is due to the hot spot, 
we calculate  the contribution of the hot spot as:
$F_{hs} = F_{max} - F_{min}$, 
where $F_{hs}$ is the flux emitted by the hot spot,
$F_{max}$  and $F_{min}$ are maximum and minimum of the observed flux 
during the night. 
They are estimated using the minimum and maximum 
magnitudes during the night (see Table~1) after correction for the observational 
errors and the scatter of the brightest and faintest points. 
To calculate the temperature, we use  the calibration for the $(B-V)$ colour 
of a black body (Table 18 in $[^{15}]$), and find. 
temperature of the hot spot 
$T= 24700 \pm 4000$~K and $T= 13350 \pm 3000$~K
for our observations on 2025-11-13 and 2025-11-14, respectively.
Our results indicate that the hot spot is probably a few $\times 1000$~K hotter 
than the value assumed in $[^{13}]$, and give a clue why they 
estimated a higher mass accretion rate. 
On 2025-11-14 the star is brighter than on 2025-11-13, and the hot spot
has a lower temperature. This can be a result that in 
brighter state the spot expands and its temperature decreases
(more details can be found in Sect. 2.3$[^{16}]$, where a model of accretion disc is 
developed).

\section{Discussion} 
\label{s.dis}

It was suggested$[^2]$ that GP Com's disc is on the lower
branch of the thermal instability curve and does not undergo 
the global outbursts followed by intervals of mass accumulation 
as occur in dwarf novae. However recently,  
one outburst was found in the  archival historical data from 1950-1951$[^{17}]$. 
This means that the outbursts are rare, probably as a result of the 
low mass accretion rate ($\approx 3 \times 10^{-12}$~M$_\odot$~yr$^{-1}$,  see Sect.\ref{s.Ma}) 
and as a
consequence long time (probably a few decades) needed for mass accumulation.

The Doppler tomography of GP~Com$[^7]$ detected a strong bright spot 
near the expected accretion stream/disc impact region,
with a faint secondary bright spot at $\sim 120$-degree 
phase-offset. The presence of a second bright spot in the accretion disc
was also found  in SDSS J120841.96+355025.2 ($P_{orb} = 52.9$~min)$[^{18}]$. 
In cataclysmic variables the hot spot must move around, 
both azimuthally and radially, as the accretion
disc and accretion rate change$[^{19}]$. 
The accretion flow originating from the inner Lagrangian point
probably has variable direction and velocity.  
In this way it strikes the accretion disc at different positions. 
Do the strong bright spot and the faint secondary bright spot (see also Sect.~\ref{s.or})
exist simultaneously or in some orbital cycles exists only one of them 
is a question that should be addressed in future work.

\vskip 0.25cm 
{\bf Conclusions:}  We report 10.6  hours photometric observations of the ultrashort period 
binary GP Com,  consisting of a white dwarf and the helium core of a former main sequence star. 
The observations are obtained with the telescopes of Rozhen National Astronomical Observatory.
We find an orbital modulation with amplitude $0.04 - 0.05$~mag in B and V bands
and variable orbital phase of the maximum. 
We estimate temperature of the bright spot $\approx 19700 \pm 3000$~K
and mass accretion rate of about $2 \times 10^{-12}$~M$_\odot$~yr$^{-1}$,
consistent with the predicted rate for a cool donor. 

\vskip 0.25cm 

{\small {\bf Acknowledgments: }
This work is part of the project KP-06-H98/8 ''Accretion flows in binary stars''
(Bulgarian National Science Fund).   
The 1.5m telescope is part of the National Roadmap for Scientific Infrastructure, 
coordinated by the Ministry of Education and Science of Bulgaria.

 \vskip 0.2cm 
{\it 
$^1$Institute of Astronomy and National Astronomical Observatory, 
Bulgarian Academy of Sciences, Tsarigradsko Shose 72, BG-1784, Sofia, Bulgaria 
  \vskip 0.1cm 
$^2$Space Research and Technology Institute, Bulgarian Academy of Sciences, 
    str. "Acad. Georgy Bonchev" bl. 1, BG-1113, Sofia 
}    
\vskip 0.2cm 
e-mail: rzamanov@nao-rozhen.org, lubav@astro.bas.bg, mminev@astro.bas.bg, \\ 
        danvasan@space.bas.bg,  f7@space.bas.bg

}

\end{document}